\documentclass[twocoloumn]{IEEEtran}
\usepackage{amsmath,amssymb,epsfig,color,flushend,algorithm,epstopdf}
\usepackage{amssymb,color,epsfig,epstopdf}

\allowdisplaybreaks

\begin{document}
\title{Secure Full-Duplex Two-Way Relaying for SWIPT}
\author{Alexander A. Okandeji, Muhammad R. A. Khandaker, Kai-Kit Wong,\\Gan Zheng, Yangyang Zhang, and Zhongbin Zheng
\thanks{\scriptsize{A. A. Okandeji, M. R. A. Khandaker and K.-K. Wong are with the Department of Electronic and Electrical Engineering, University College London, WC1E 7JE, United Kingdom (e-mail: $\rm alexander.okandeji.13@ucl.ac.uk$).
\par G. Zheng is with Wolfson School of Mechanical, Electrical and Manufacturing Engineering, Loughborough University, United Kingdom.
\par Y. Zhang is with Kuang-Chi Institute of Advanced Technology, Shenzhen, China.
\par Z. Zheng is with East China Institute of Telecommunications, China Academy of Information and Communications Technology, Shanghai, China.
\par This work was supported by the Presidential Special Scholarship Scheme for Innovation and Development (PRESSID), Federal Republic of Nigeria.}}
}
\maketitle
\begin{abstract}
This letter studies bi-directional secure information exchange in a simultaneous wireless information and power transfer (SWIPT) system enabled by a full-duplex (FD) multiple-input multiple-output (MIMO) amplify-and-forward (AF) relay. The AF relay injects artificial noise (AN) in order to confuse the eavesdropper. Specifically, we assume a zeroforcing (ZF) solution constraint to eliminate the residual self-interference (RSI). As a consequence, we address the optimal joint design of the ZF matrix and the AN covariance matrix at the relay node as well as the transmit power at the sources. We propose an alternating algorithm utilizing semi-definite programming (SDP) technique and one-dimensional searching to achieve the optimal solution. Simulation results are provided to demonstrate the effectiveness of the proposed algorithm.
\end{abstract}

\section{Introduction}
Recently, simultaneous wireless information and power transfer (SWIPT) \cite{mimo, ruhuul2, jrnl_rob_sec_swipt, Alex, liao, ruhuul2a, jrnl_2way_eh_relay, ruhuul, QOS}, full-duplex (FD) enabled bi-directional wireless communications \cite{Alex, off-the-shelf, Alex_2} as well as physical-layer (PHY) security \cite{eaves} have each been a major research area and also led to efforts investigating the combination of these technologies. To name a few, for example, FD SWIPT has been considered in \cite{Alex_3}. Also, PHY security in FD systems was addressed in \cite{ying,yaping}.
In contrast to previous work, our main contribution is the study of the integration of all three and the joint optimization of the sources' transmit power, the artificial noise (AN) covariance and the two-way relay beamforming matrix to maximize the secrecy sum-rate for SWIPT with a FD multiple-input multiple-output (MIMO) amplify-and-forward (AF) relay employing power splitter (PS). Specifically, the total transmit power is minimized while guaranteeing the signal to interference and noise ratio (SINR) constraints at the two legitimate users as well as the eavesdropper and the energy harvesting constraint at the relay.

{\em Notations}--We use ${\sf X}\in\mathbb{C}^{M \times N}$ to represent a complex $M \times N$ matrix. Also, $(\cdot)^{\dagger}$ denotes the conjugate transpose, $\mathrm{trace}(\cdot)$ is the trace operation, and $\|\cdot\|$ denotes the Frobenius norm. In addition, $| \cdot|$ returns the absolute value of a scalar, and ${\sf X}\succeq {\bf 0}$ denotes that the Hermitian matrix ${\sf X}$ is positive semidefinite. The expectation operator is denoted by $\mathbb{E}\{\cdot\}.$

\section{System Model}
We consider SWIPT in a three-node MIMO relay network with sources $\rm S_{A}$ and $\rm S_{B},$ consisting of one transmit and receive antenna for information transmission and reception, respectively, exchanging confidential information with the aid of a multiantenna AF relay $\sf R,$ in the presence of a single antenna eavesdropper $\sf {E}$. The relay harvests energy to complete the information exchange as also assumes in \cite{jrnl_rob_sec_swipt}. We assume that: i) $\rm S_{A}$, $\rm S_{B}$ and $\sf R$ all operate in FD mode, ii) there is no direct link between $\rm S_{A}$ and $\rm S_{B}$, and iii) the source nodes are not aware of any eavesdropper thus, no direct link exist between the source nodes and the eavesdropper \cite{sum}. The relay however, is aware of the eavesdropper. As a result, the relay injects AN signals to confuse the eavesdropper.

In the first phase, the relay receives confidential information from $\rm S_{A}$ and $\rm S_{B}$, while in the next phase, $\sf R$ amplifies and forwards the processed information to both sources with the AN signal being superimposed to jam the eavesdropper \cite{sum}. The harvested energy at the relay is used to complete the bi-directional information exchange between the source nodes. Using the transmit power ${P}_A$ and $ {P}_B,$ respectively, $\rm S_{A}$ and $\rm S_{B}$ transmit their confidential messages simultaneously to $\sf{R}.$ On the other hand, $\sf{R}$ employs linear processing with amplification matrix ${\mathbf{W}}$ to process the received signal and broadcasts the processed signal to the nodes with harvested power ${\rm U}$.

The antennas at ${\sf R}$ are separated for transmission and reception with $M_T$ transmit antennas and $M_R$ receive antennas. Also, we denote $\mathbf{h}_{XR} \in \mathbb{C}^{M_R \times 1}$ and $\mathbf{h}_{RX} \in \mathbb{C}^{M_T \times 1}$ to, respectively, represent the directional channel vectors between the source node {\sf X} $\in\{A, B\}$ and ${\sf R}$. Similarly, we use $\mathbf{h}_{RE}$ to denote the channel between ${\sf E}$ and $\sf{R}.$

To achieve FD communication, self-interference (SI) must be significantly suppressed, as total cancellation is not possible as a result of imperfect channel estimation \cite{Alex_3}. Therefore, we adopt the use of existing SI cancellation mechanisms (e.g., antenna isolation, digital and analog cancellation, etc.), to reduce the effect of SI. For convenience, we denote $h_{AA},$ $h_{BB},$ and $\mathbf{H}_{RR}\in \mathbb{C}^{M_R \times M_T}$ as the residual SI (RSI) channels at the respective nodes \cite{Alex_3}. Also, the RSI channel is represented as a Gaussian distribution random variable with zero mean and variance $\sigma^2_X,$ for ${\sf X} \in \{A, B, R\}$ \cite{Alex_3}. Furthermore, the relay, assumed to be equipped with a PS device, coordinates information decoding and energy harvesting. Specifically, the relay splits the received signal power such that a $\rho \in (0,1)$ portion of the received signal power is fed to the information receiver (IR) and the remaining $(1-\rho)$ portion of the power is fed to the energy receiver (ER) at the relay.

\subsection{Signal Model}
The received signal ${\mathbf{y}}_r[n]$ and the transmit signal $\mathbf{x}_R[n]$ at ${\sf R}$ at time instant $n,$ can be written, respectively, as
\begin{align}
{\mathbf{y}}_r[n] \!&=\!  \mathbf{h}_{AR}s_A[n] + \mathbf{h}_{BR}s_B[n]+  \mathbf{H}_{RR}{\mathbf{x}}_R[n] + \mathbf{n}_R[n],\\
{\mathbf{x}}_R[n] \!&=\! {\mathbf{W}}{\mathbf{y}}^{IR}_R [n-\tau] + \mathbf{z}[n], \label{beam}
\end{align}
where $\tau$ is the processing delay to implement FD operation and assumed short enough to be neglected as far as the achievable rate computation is concerned, $\mathbf{n}_R \backsim \mathcal{CN} (\mathbf{0}, \sigma^2_R\mathbf{I})$ is the additive white Gaussain noise (AWGN) at ${\sf R}$, $\mathbf{z}[n] \backsim \mathcal{CN} (\mathbf{0}, \mathbf{Q}),$ with $\mathbf{Q} \succeq \mathbf{0}$, is the AN used for interfering ${\sf E},$ and $\mathbf{y}^{IR}_R[n]$ is the signal split to the IR at ${\sf R}$ given by
\begin{equation}
\mathbf{y}^{IR}_R[n] = \sqrt{\rho} \Big(\mathbf{h}_{AR}s_A[n] + \mathbf{h}_{BR}s_B[n] + \mathbf{H}_{RR}{\mathbf{x}}_R[n]
+ \mathbf{n}_R[n]\Big).
 \label{receive_relay1}
\end{equation}
Thus, the signal transmitted by ${\sf R}$ can then be expressed as
\begin{multline}
{\mathbf{x}}_R[n] = \sqrt{\rho}  {\mathbf{W}}   \Big(\mathbf{h}_{AR}s_A[n-\tau] +  \mathbf{h}_{BR}s_B[n-\tau]\\
 + \mathbf{H}_{RR} {\mathbf{x}}_R[n-\tau] + \mathbf{n}_R[n-\tau]\Big)\! +\! \mathbf{z}[n-\tau]. \label{beam2}
\end{multline}
As shown in \cite{delay}, the relay output can be further written as
\begin{multline}
{\mathbf{x}}_R[n] = {\mathbf{W}} \sum^{\infty}_{j = 0} (\mathbf{H}_{RR} {\mathbf{W}})^j \Big[ \sqrt{\rho} (\mathbf{h}_{AR}s_A[n-j\tau-\tau]\\
+ \mathbf{h}_{BR}s_B[n-j\tau-\tau] + \mathbf{n}_R[n-j\tau-\tau])\Big]\\
+ \mathbf{z}[n-j\tau-\tau], \label{beam3}
\end{multline}
where $j$ denotes the index of the delayed symbols. We define the covariance matrix of (\ref{beam3})  as
\begin{multline}
\mathbb{E}[{\mathbf{x}}_R {\mathbf{x}}^{\dagger}_R] \!\!=\!\!  \rho \Big[\!P_A \! {\mathbf{W}}\!\sum^{\infty}_{j=0} (\mathbf{H}_{RR} {\mathbf{W}})^j \mathbf{h}_{AR}\mathbf{h}_{AR}^\dagger (( \mathbf{H}_{RR} {\mathbf{W}} )^j)^\dagger {\mathbf{W}}^\dagger \\
+ P_B  {\mathbf{W}}\sum^{\infty}_{j=0} (\mathbf{H}_{RR} {\mathbf{W}})^j \mathbf{h}_{BR}\mathbf{h}_{BR}^\dagger (( \mathbf{H}_{RR} {\mathbf{W}} )^j)^\dagger {\mathbf{W}}^\dagger \\
+{\mathbf{W}} \sum^{\infty}_{j=0}( \mathbf{H}_{RR} {\mathbf{W}} {\mathbf{W}}^\dagger \mathbf{H}_{RR}^\dagger )^j {\mathbf{W}}^\dagger \Big] + \mathbf{Q}. \label{cov^}
\end{multline}
Clearly, the relay's transmit covariance is indeed a complicated function of ${\mathbf{W}}.$ In this letter, we adopt the zeroforcing (ZF) solution constraint to cancel the RSI from the relay output to the relay input via the optimization of ${\mathbf{W}}$ \cite{Alex_3}. In particular, the ZF constraints may take the following forms \cite{sum}
\begin{align}
\mathrm{\mathbf{W}} \mathbf{H}_{RR} = {\bf 0}, & ~~\mbox{if }M_R > M_T,\label{ZF1}\\
\mathrm{\mathbf{H}}_{RR} \mathbf{W} = {\bf 0}, & ~~\mbox{if }M_T > M_R.\label{ZF2}
\end{align}
For convenience, we only consider the case ${M}_T > {M}_R$ as the other case can be handled similarly. Thus, (\ref{beam3}) becomes
\begin{equation}
{\mathbf{x}}_R[n]\!\! =\!\! \sqrt{\rho} \mathbf{W}\Big[\!  \mathbf{h}_{AR}s_A[n-\tau] + \mathbf{h}_{BR}s_B[n-\tau] + \mathbf{n}_R[n-\tau]\Big]\!+ \mathbf{z}[n],\nonumber
  \label{beam4}
\end{equation}
with the relay output power expressed as
 \begin{multline}
P_R = {\rm trace} (\mathbb E[\bf x_R \bf x_R^{\dagger}]) \\=  \rho  \Big [ P_A{ \|\mathbf{W} \mathbf{h}_{AR}\|^2}  + P_B \| \mathbf{W} \mathbf{h}_{BR}\|^2 + \mathrm{trace} ( \mathbf{W}\mathbf{W}^{\dagger}) \Big] \\
  \qquad\qquad + \mathrm{trace} \mathbf{(Q)}.
\end{multline}
In the second time slot after cancelling the SI signal $s_A[n - \tau]$, the received signal at $\rm S_A$ is given as
\begin{multline}
y_{SA}[n] =  \sqrt{\rho}\Big (\mathbf{h}^{\dagger}_{RA} {\mathbf{W}} \mathbf{h}_{BR} s_B[n-\tau]
 + \mathbf{h}^{\dagger}_{RA}  {\mathbf{W}} \mathbf{n}_R[n]\Big) \\
 + \mathbf{h}^{\dagger}_{RA} \mathbf{z}[n] + {h}_{AA}s_A[n] +  n_A[n],\label{signode2}
\end{multline}
where $n_A[n]$ is the AWGN at source node ${\rm S}_{\rm A}$. From this, we can work out the rates at $\rm S_A$ and $\rm S_B$ as
\begin{equation}
R_X=\log_2(1+\Gamma_X),~~\mbox{for } { X}\in\{A,B\},
\end{equation}
where
\begin{align}
\Gamma_A&=\frac{\rho P_B |\mathbf{h}^{\dagger}_{RA} \mathbf{W} \mathbf{h}_{BR}|^2}{ \rho \sigma^2_R \|\mathbf{h}^{\dagger}_{RA} \mathbf{W}\|^2 \! +\! P_A| {h_{AA}}|^2\! + \!\mathbf{h}^{\dagger}_{RA}\mathbf{Q}\mathbf{h}_{RA} \!+\! 1}, \label{13}\\
\Gamma_B&=\frac{\rho P_A |\mathbf{h}^{\dagger}_{RB} \mathbf{W} \mathbf{h}_{AR}|^2}{ \rho \sigma^2_R \|\mathbf{h}^{\dagger}_{RB} \mathbf{W}\|^2 \! +\! P_B| {h_{BB}}|^2 \!+\! \mathbf{h}^{\dagger}_{RB}\mathbf{Q}\mathbf{h}_{RB}\! +\! 1}.\label{14}
\end{align}
The signal received at {$\sf E$} can be expressed as
\begin{multline}
\gamma_E[n]\! =\!\! \sqrt{\rho} \Big(  \mathbf{h}^{\dagger}_{RE} {\mathbf{W}}\mathbf{h}_{AR}s_A[n-\tau]\! +\!  \mathbf{h}^{\dagger}_{RE} {\mathbf{W}} \mathbf{h}_{BR}s_B[n-\tau]\\
+ \mathbf{h}^{\dagger}_{RE}\mathbf{W} \mathbf{n}_R \! \Big) +\! \mathbf{h}^{\dagger}_{RE} \mathbf{z}[n]\! +\! n_E,
\end{multline}
where $n_E$ is the AWGN at {$\sf E$}.
Also, the achievable sum-rate at {\sf E} is upper bounded as
$R_E  = \log_2 (1+\Gamma_E)$ \cite{sum},
where
\begin{equation}
\Gamma_E=\frac{ \rho P_A |\mathbf{h}^{\dagger}_{RE} {\mathbf{W}}\mathbf{h}_{AR}|^2\! +\! \rho P_B |\mathbf{h}^{\dagger}_{RE} {\mathbf{W}} \mathbf{h}_{BR}|^2}{\rho \sigma^2_R \|\mathbf{h}^{\dagger}_{RE}\mathbf{W}\|^2\! +\! \mathbf{h}^{\dagger}_{RE}\mathbf{Q} \mathbf{h}_{RE}\! +\! 1 }.
\end{equation}
The achievable secrecy sum-rate is then defined as \cite{sum}
\begin{equation}
R_{sec} = [R_A + R_B - R_E]^+, \label{R_sec}
\end{equation}
where $[x]^+$ represents $\max (x,0)$.  
Meanwhile, the signal split to the ER at {$\sf R$} is given by
\begin{equation}
\mathbf{y}^{ER}_R[n] \!\! = \!\! \sqrt{1-\rho} \Big(\!\mathbf{h}_{AR}s_A[n]\!\! +\!\! \mathbf{h}_{BR}s_B[n]
 \!\! +\!\! \mathbf{H}_{RR}\mathrm{\mathbf{x}}_R[n]\!\! +\!\! \mathbf{n}_R[n]\!\Big). \nonumber
\end{equation}
The harvested energy at the relay is thus given as \cite{Alex_3}
\begin{eqnarray}
\rm U = \beta({1-\rho})(|\mathbf{h}_{AR}|^2P_A + |\mathbf{h}_{BR}|^2P_B + \mathrm{\bar{E}} + \sigma^2_R M_R),
\end{eqnarray}
in which $\mathrm{\bar{E}} = \mathbb{E}[\mathrm{\mathbf{x}}_R \mathrm{\mathbf{x}}^{\dagger}_R]$ and $\beta$ denotes the energy conversion efficiency of the ER at the relay which is assumed unity.

\subsection{Problem Statement}
Due to the inherent SI at each FD node, the source nodes may not use the maximum available transmit power in order not to increase the level of SI. Thus, there is a need to transmit at optimum values. Furthermore, it is known that optimal values of system parameters guarantees that the secrecy rate is as large as possible \cite{sum}. Thus, in this letter, our aim is to maximize the secrecy sum-rate for SWIPT
by ensuring system parameters are optimal.
We achieve this by jointly optimizing the transmit power at the source nodes ($P_A, P_B$), the relaying matrix ($\mathbf{W}$) and the AN covariance matrix ($\mathbf{Q}$) at the relay. Thus, we have
\begin{equation}\label{main_prob1}
\begin{aligned}
&\min_{\rho \in (0,1), {\bf \mathbf{W}},\mathbf{Q}\succeq \mathbf{0}\atop 0 < P_A \leq P_{\mathrm{max}}, 0 < P_B \leq P_{\mathrm{max}}} P_A + P_B + P_R~~
\mbox{s.t.}\\
& \left\{\begin{aligned}
\Gamma_A &\geq \gamma_A,\\
\Gamma_B &\geq \gamma_B,\\
\Gamma_E & \leq \gamma_E,\\
({1-\rho})(|\mathbf{h}_{AR}|^2P_A \! +\! |\mathbf{h}_{BR}|^2P_B \! +\! \mathrm{\bar{E}} +\! \sigma^2_R M_R) & \geq \bar{\rm U},\\
  \mathbf{H}_{RR} \mathrm{\mathbf{W}}& = {\bf 0}.
\end{aligned}\right.
\end{aligned}
\end{equation}
As (\ref{main_prob1}) is nonconvex, we solve (\ref{main_prob1}) in an alternating manner.
\section{Proposed Scheme}
\subsection{Optimization of $\mathbf{W}$ and $\mathbf{Q}$ at the Relay}
Here, we study the optimal beamforming matrix and the AN covariance matrix assuming the source power ($P_A, P_B$) and the PS ratio ($\rho$) all being fixed. For convenience, we define $\mathrm{\mathbf{W}}$ = $\mathbf{N}_t \mathbf{V},$ where $\mathbf{N}_t\in \mathbb{C}^{M_T \times M_T}$ represents the null space of $\mathrm{\mathbf{H}}_{RR},$ and $\mathbf{V} \in \mathbb{C}^{M_T \times M_T}$ is the new optimization variable. As a consequence, the optimization of $\mathbf{W}$ reduces to optimizing $\mathbf{V}.$ 
Hence, we remove the ZF constraint in (\ref{main_prob1}) and obtain the equivalent optimization problem:
\begin{equation}\label{main_prob1a}
\begin{aligned}
&\min_{{\bf{V}, \mathbf{Q}\succeq \mathbf{0}}} P_R~~\mbox{s.t.}\\
&\left\{\begin{aligned}
\Gamma_A &\geq \gamma_A,\\
\Gamma_B &\geq \gamma_B,\\
\Gamma_E &\leq \gamma_E,\\
({1-\rho})(|\mathbf{h}_{AR}|^2P_A\!\! +\! |\mathbf{h}_{BR}|^2P_B \!+\! \mathrm{\bar{E}}\! +\! \sigma^2_R M_R) &\geq \bar{\rm U}.
\end{aligned}\right.
\end{aligned}
\end{equation}
Problem (\ref{main_prob1a}) is a nonconvex problem due to the coupled optimization variables in the constraints. However, by rearranging the terms in the constraints, (\ref{main_prob1a}) can be re-expressed as
\begin{subequations}\label{main_prob1w}
\begin{eqnarray}
\!\!\!& &\!\!\!  \min_{{\mathbf{\Sigma}, \mathbf{Q}\succeq \mathbf{0}}} P_R ~~\mbox{s.t.} \\
\!\!\!& &\!\!\! \frac{1} {\gamma_A} P_B C_{rA} \mathbf{h}_{BR}^{\dagger} \mathbf{\Sigma} \mathbf{h}_{BR} \!-\! \sigma^2_R C_{Nt} \mathbf{h}_{RA}^{\dagger} \mathbf{\Sigma} \mathbf{h}_{RA} \nonumber\\
\!\!\!& &\!\!\! \qquad\qquad \geq  \frac{1}{\rho}(P_A|h_{AA}|^2 \!\! +\!\! \mathbf{h}_{RA}^{\dagger}\mathbf{Q}\mathbf{h}_{RA}+1), \\
\!\!\!& &\!\!\! \frac{1}{\gamma_B} P_A C_{rB} \mathbf{h}_{AR}^{\dagger} \mathbf{\Sigma} \mathbf{h}_{AR} \! - \! \sigma^2_R C_{Nt} \mathbf{h}_{RB}^{\dagger} \mathbf{\Sigma}\mathbf{h}_{RB} \nonumber\\
\!\!\!& &\!\!\! \qquad\qquad \geq \frac{1}{\rho}(P_B|h_{BB}|^2 \!\!+\!\! \mathbf{h}_{RB}^{\dagger} \mathbf{Q}\mathbf{h}_{RB}\!\!+\!\! 1), \\
\!\!\!& &\!\!\! \frac{1}{\gamma_E}\Big[ \! P_A C_{rE} \mathbf{h}_{AR}^{\dagger} {\mathbf{\Sigma}}\mathbf{h}_{AR} \!+\! P_B  C_{rE} \mathbf{h}_{BR}^{\dagger} {\mathbf{\Sigma}} \mathbf{h}_{BR}\Big] \! \nonumber\\
\!\!\!& &\!\!\! \! - \sigma^2_R C_{Nt} \mathbf{h}_{RE}^{\dagger}\mathbf{\Sigma} \mathbf{h}_{RE}
 \leq \frac{1}{\rho}\left(\mathbf{h}_{RE}^{\dagger} \mathbf{Q} \mathbf{h}_{RE}\! +\! 1\right), \\
\!\!\!& &\!\!\! |\mathbf{h}_{AR}|^2P_A + |\mathbf{h}_{BR}|^2P_B \!+ \mathrm{\bar{E}} \!\geq\!\frac{\rm U}{(1-\rho)}-\! \sigma^2_R M_R,
\label{main_prob1w^*}
\end{eqnarray}
\end{subequations}
where $\mathbf{\Sigma} = \mathbf{V}\mathbf{V}^{\dagger}, C_{rA}= \|\mathbf{N}_t \mathbf{h}_{RA} \|^2, C_{Nt} = \mathrm{trace} (\mathbf{N}_t\mathbf{N}^{\dagger}_t), C_{rB}= \|\mathbf{N}_t \mathbf{h}_{RB} \|^2$ and $C_{rE}= \|\mathbf{N}_t \mathbf{h}_{RE} \|^2. $
Problem (\ref{main_prob1w}) can be efficiently solved by existing solvers such as CVX \cite{boyd}.
Once the optimal $\mathbf{\Sigma}$ is obtained, optimal $\mathbf{V}$ can be constructed through matrix decomposition.

\subsection{Optimization of the PS Coefficient $(\rho)$}
For fixed values of the relay beamforming matrix ($\mathbf{W}$), AN covariance ($ \mathbf{Q}$) and for given values of the transmit power ($P_A, P_B$) at the sources, (\ref{main_prob1}) can be reformulated as
\begin{subequations}\label{main_prob1^*}
\begin{eqnarray}
\!\!\!& &\!\!\! \min_{\rho \in (0,1)}  \quad  P_A + P_B + P_R \quad \quad {\rm s.t.} \\
\!\!\!& &\!\!\!    \frac{\rho P_B |\mathbf{h}^{\dagger}_{RA} \mathbf{W} \mathbf{h}_{BR}|^2}{ \rho \sigma^2_R \|\mathbf{h}^{\dagger}_{RA} \mathbf{W}\|^2 \!\!+\!\! P_A| {h_{AA}}|^2 \!\! +\!\! \mathbf{h}^{\dagger}_{RA}\mathbf{Q}\mathbf{h}_{RA}\!\! +\!\! 1} \!\! \geq\!\! \gamma_A, \label{main_opt_a} \\
\!\!\!& &\!\!\!   \frac{\rho P_A |\mathbf{h}^{\dagger}_{RB} \mathbf{W} \mathbf{h}_{AR}|^2}{ \rho \sigma^2_R \|\mathbf{h}^{\dagger}_{RB} \mathbf{W}\|^2\!\!  +\!\! P_B| {h_{BB}}|^2 \!\!+\!\! \mathbf{h}^{\dagger}_{RB}\mathbf{Q}\mathbf{h}_{RB} \!\!+\!\! 1} \!\!\geq\!\! \gamma_B, \label{main_opt_b} \\
\!\!\!& &\!\!\!  \frac{ \rho P_A |\mathbf{h}^{\dagger}_{RE} {\mathbf{W}}\mathbf{h}_{AR}|^2 + \rho P_B |\mathbf{h}^{\dagger}_{RE} {\mathbf{W}} \mathbf{h}_{BR}|^2}{\rho \sigma^2_R \|\mathbf{h}^{\dagger}_{RE}\mathbf{W}\|^2 + \mathbf{h}^{\dagger}_{RE}\mathbf{Q} \mathbf{h}_{RE} + 1 } \leq \gamma_E, \label{main_opt_c} \\
\!\!\!& &\!\!\! ({1-\rho})(|\mathbf{h}_{AR}|^2P_A \!+ \!|\mathbf{h}_{BR}|^2P_B \!+\! \mathrm{\bar{E}} \!+\!\! \sigma^2_R M_R)\!\!\geq\! \bar{\rm U}\!, \label{main_opt_d}
\end{eqnarray}
\end{subequations}
which can be expressed in a form solvable by existing solvers by rearranging the terms in the constraints as
\begin{subequations}\label{main_prob1^**}
\begin{eqnarray}
 \!\!\!& &\!\!\! \min_{\rho \in \{0,1\}} P_A + P_B + P_R  \quad \quad {\rm s.t.}  \\
   \!\!\!& &\!\!\! \frac{1}{\gamma_A} \rho P_B C_{rA}  \mathbf{h}_{BR}^{\dagger} \mathbf{\Sigma} \mathbf{h}_{BR} - \rho \sigma^2_R C_{Nt} \mathbf{h}_{RA}^{\dagger} \mathbf{\Sigma} \mathbf{h}_{RA} \nonumber\\
 \!\!\!& &\!\!\! \qquad\qquad\qquad \geq P_A| {h_{AA}}|^2 + \mathbf{h}_{RA}^{\dagger}\mathbf{Q}\mathbf{h}_{RA} + 1,  \label{main_opt_a*} \\
 \!\!\!& &\!\!\!  \frac{1}{\gamma_B} \rho P_A C_{rB}  \mathbf{h}_{AR}^{\dagger} \mathbf{\Sigma} \mathbf{h}_{AR} - \rho \sigma^2_R C_{Nt} \mathbf{h}_{RB}^{\dagger} \mathbf{\Sigma} \mathbf{h}_{RB} \nonumber\\
\!\!\!& &\!\!\! \qquad\qquad\qquad \geq P_B| {h_{BB}}|^2 + \mathbf{h}_{RB}^{\dagger} \mathbf{Q}\mathbf{h}_{RB} + 1,  \label{main_opt_B*} \\
 \!\!\!& &\!\!\!  \frac{1}{\gamma_E} \Big[  P_A C_{rE}\mathbf{h}_{AR}^{\dagger} {\mathbf{\Sigma}}\mathbf{h}_{AR} \! + \! P_B C_{rE}\mathbf{h}_{BR}^{\dagger} {\mathbf{\Sigma}} \mathbf{h}_{BR}\Big] \nonumber\\
\!\!\!& &\!\!\! - \sigma^2_R C_{Nt} \mathbf{h}_{RE}^{\dagger} \mathbf{\Sigma} \mathbf{h}_{RE}
  \leq \frac{1}{\rho}\left(\mathbf{h}_{RE}^{\dagger} \mathbf{Q} \mathbf{h}_{RE} + 1\right),  \label{main_opt_c*} \\
  \!\!\!& &\!\!\! ({1-\rho})(|\mathbf{h}_{AR}|^2P_A \!+\! |\mathbf{h}_{BR}|^2P_B\! +\! \mathrm{\bar{E}}\! + \! \sigma^2_R M_R)\!\! \geq \!\! \bar{\rm U}\!. \label{main_opt_d*}
\end{eqnarray}
\end{subequations}
\subsection{Optimization of the Source Power $(P_A, P_B)$}
For given values of the relay beamforming matrix ($\mathbf{W}$), AN covariance matrix ($\mathbf{Q}$) and the relay PS ratio, problem (\ref{main_prob1}) can be written as
\begin{subequations} \label{main_prob11}
\begin{eqnarray}
\!\!\!& &\!\!\! \min_{{P_A, P_B}} \quad  P_A + P_B + P_R \quad \quad {\rm s.t.} \nonumber\\
    \!\!\!& &\!\!\!    \frac{\rho P_B |\mathbf{h}^{\dagger}_{RA} \mathbf{W} \mathbf{h}_{BR}|^2}{ \rho \sigma^2_R \|\mathbf{h}^{\dagger}_{RA} \mathbf{W}\|^2 \!+\! P_A| {h_{AA}}|^2 \!\!+\!\! \mathbf{h}^{\dagger}_{RA}\mathbf{Q}\mathbf{h}_{RA} \!+\! 1}\!\!\! \geq \!\! \gamma_A \!, \label{m_pa}\\
 \!\!\!& &\!\!\!   \frac{\rho P_A |\mathbf{h}^{\dagger}_{RB} \mathbf{W} \mathbf{h}_{AR}|^2}{ \rho \sigma^2_R \|\mathbf{h}^{\dagger}_{RB} \mathbf{W}\|^2 \! +\! P_B| {h_{BB}}|^2\!\! +\!\! \mathbf{h}^{\dagger}_{RB}\mathbf{Q}\mathbf{h}_{RB} \!\! +\!\! 1}\!\! \geq \!\! \gamma_B\!,  \label{m_pb}\\
 \!\!\!& &\!\!\!  \frac{ \rho P_A |\mathbf{h}^{\dagger}_{RE} {\mathbf{W}}\mathbf{h}_{AR}|^2 + \rho P_B |\mathbf{h}^{\dagger}_{RE} {\mathbf{W}} \mathbf{h}_{BR}|^2}{\rho \sigma^2_R \|\mathbf{h}^{\dagger}_{RE}\mathbf{W}\|^2 + \mathbf{h}^{\dagger}_{RE}\mathbf{Q} \mathbf{h}_{RE} + 1 } \leq \gamma_E\!,  \label{m_pc} \\
  \!\!\!& &\!\!\! ({1-\rho})(|\mathbf{h}_{AR}|^2P_A \!+\! |\mathbf{h}_{BR}|^2P_B\! +\! \mathrm{\bar{E}} \!+ \! \sigma^2_R M_R) \!\!\geq \!\! \bar{\rm U}\!,  \label{m_pd}\\
\!\!\!& &\!\!\! 0 < P_A \leq P_{\mathrm{max}}, \quad 0 < P_B \leq P_{\mathrm{max}}\label{m_pf}.
\end{eqnarray}
\end{subequations}
It is worth noting that full-duplexity in communication systems is preceded by successful SI cancellation. In our model, the source nodes are equipped with a single transmitter-receiver pair for signal transmission and reception, respectively. As a result, it is impossible to cancel the SI in the spatial domain \cite{Alex_3}. The relay, in contrast, equipped with at least two transmitter-receiver pairs, can cancel the generated SI in the spatial domain. We proceed to investigate the optimal power solution ($P_A, P_B$) assuming $\mathbf{W},$ $\mathbf{Q}$ and $\rho$ all being fixed. Evidently, it is easy to check that at the optimum, at least one source will be transmitting with maximum power \cite{Alex_3} i.e., $P_A = P_{\mathrm{max}}$ or $P_B = P_{\mathrm{max}}.$ As a consequence, we can relax (\ref{main_prob11}) into two sub-problems with: (i) $P_A = P_{\mathrm{max}}$, (ii) $P_B = P_{\mathrm{max}}.$ Considering the symmetric nature of case (i) and case (ii), we study case (i) as an example and solve problem (\ref{main_prob11}) analytically. Problem (\ref{main_prob11}) is thus reformulated as
 \begin{subequations} \label{main_pr1^}
\begin{eqnarray}
\!\!\!& &\!\!\! \min_{{P_B}} \quad P_B + {\bar P}_R  \quad \quad  {\rm s.t.} \\
    \!\!\!& &\!\!\!    \frac{\rho P_B |\mathbf{h}^{\dagger}_{RA} \mathbf{W} \mathbf{h}_{BR}|^2}{ \rho \sigma^2_R \|\mathbf{h}^{\dagger}_{RA} \mathbf{W}\|^2 \!\!+\!\! P_{\mathrm{max}}\!| {h_{AA}}|^2 \!\!+\!\! \mathbf{h}^{\dagger}_{RA}\mathbf{Q}\mathbf{h}_{RA} \!\!+\!\! 1} \!\!\geq\!\! \gamma_A\!,  \label{m_pa^}\\
 \!\!\!& &\!\!\!   \frac{\rho P_{\mathrm{max}} |\mathbf{h}^{\dagger}_{RB} \mathbf{W} \mathbf{h}_{AR}|^2}{ \rho \sigma^2_R \|\mathbf{h}^{\dagger}_{RB} \mathbf{W}\|^2  \!\!+\!\! P_B| {h_{BB}}|^2\!\! +\!\! \mathbf{h}^{\dagger}_{RB}\mathbf{Q}\mathbf{h}_{RB} \!+ \!1} \!\!\geq\!\! \gamma_B\!,  \label{m_pb^}\\
 \!\!\!& &\!\!\!  \frac{ \rho P_{\mathrm{max}}\! |\mathbf{h}^{\dagger}_{RE} {\mathbf{W}}\mathbf{h}_{AR}|^2 + \rho P_B |\mathbf{h}^{\dagger}_{RE} {\mathbf{W}} \mathbf{h}_{BR}|^2}{\rho \sigma^2_R \|\mathbf{h}^{\dagger}_{RE}\mathbf{W}\|^2 \!+\! \mathbf{h}^{\dagger}_{RE}\mathbf{Q} \mathbf{h}_{RE}\! +\! 1 } \!\!\leq\! \! \gamma_E\!,  \label{m_pc^} \\
  \!\!\!& &\!\!\! ({1-\rho})(|\mathbf{h}_{AR}|^2\! P_{\mathrm{max}}\!\! +\! |\mathbf{h}_{BR}|^2P_B \! +\!\! \mathrm{\bar{E}} \!+\! \sigma^2_R M_R\!)\!\! \geq \!\! \bar{\rm U}\!,  \label{m_pd^}\\
\!\!\!& &\!\!\! 0 < P_B \leq P_{\mathrm{max}},
\end{eqnarray}
\end{subequations}
where ${\bar P}_R \!\! = \!\! \rho \Big [ P_{\mathrm{max}} \|\mathbf{W} \mathbf{h}_{AR}\|^2  + P_B \| \mathbf{W} \mathbf{h}_{BR}\|^2 + \mathrm{trace} (\mathbf{W}\mathbf{W}^{\dagger}) \Big] + \mathrm{trace} \mathbf{(Q)}$. Since $0 < P_B \leq P_{\mathrm{max}},$ we can obtain the feasible range [$P^{\mathrm{min}}_B, P^{\mathrm{max}}_B$] for $P_B$. Also, the constraints in (\ref{main_pr1^}) can be analysed with respect to $P_B$:\\
1) A continuous increase in $P_B$ should guarantee that (\ref{m_pa^}) remains satisfied. As a consequence, we can set the minimum of $P_B$ as
$P^{\mathrm{min}}_B = \frac{ \gamma_A(\rho \sigma^2_R \|\mathbf{h}^{\dagger}_{RA} \mathbf{W}\|^2 + P_{A}| {h_{AA}}|^2 + \mathbf{h}^{\dagger}_{RA}\mathbf{Q}\mathbf{h}_{RA} + 1)}{\rho |\mathbf{h}^{\dagger}_{RA} \mathbf{W} \mathbf{h}_{BR}|^2}.$\\
2) Constraint (\ref{m_pb^}) is a decreasing function of $P_B$. Therefore, the maximum $P_B$ satisfying (\ref{m_pb^}) to equality is defined as
$  P^{\mathrm{max}}_B = \frac{{\rho P_{A} |\mathbf{h}^{\dagger}_{RB} \mathbf{W} \mathbf{h}_{AR}|^2} - \gamma_B( \rho \sigma^2_R \|\mathbf{h}^{\dagger}_{RB} \mathbf{W}\|^2 + \mathbf{h}^{\dagger}_{RB}\mathbf{Q}\mathbf{h}_{RB} + 1 )}{\gamma_B | {h_{BB}}|^2}. $\\
3) An upper bound of the eavesdropping constraint in (\ref{m_pc^}) is satisfied when $P_B \le P^{\mathrm{max}}_B.$\\
4) A lower bound of the energy harvesting constraint in (\ref{m_pd^}) is guaranteed to be satisfied when $P_B \ge P^{\mathrm{min}}_B.$ \\
The optimal $P^*_B$ is chosen between $ P^{\mathrm{min}}_B$ and $  P^{\mathrm{max}}_B$ which satisfies (\ref{m_pa^})--(\ref{m_pd^}). Accordingly, to obtain the optimal $P_B,$ we perform a 1-D search over $P_B$ starting from $ P^{\mathrm{min}}_B$  until $ P^{\mathrm{max}}_B$ is reached to find a feasible solution to problem (\ref{main_pr1^}). Clearly, if $ P^{\mathrm{min}}_B >  P^{\mathrm{max}}_B$ then (\ref{main_pr1^}) becomes infeasible. 

\section{Simulation Results}
In this section, we present numerical results to investigate the performance of the proposed scheme through computer simulations. We consider a Rayleigh flat fading channel. The results are averaged over $1000$ independent realizations and SINR at node A, node B and the eavesdropper is given, respectively, as $\gamma_A = -5 ({\rm dB})$, $\gamma_B = -5 ({\rm dB})$, $\gamma_E =-15 ({\rm dB})$. We also assume that 60\% of the SI at node A and node B has been eliminated through digital cancellation \cite{Alex_3}.
\begin{figure}[ht]
\centering
\includegraphics*[width=8cm]{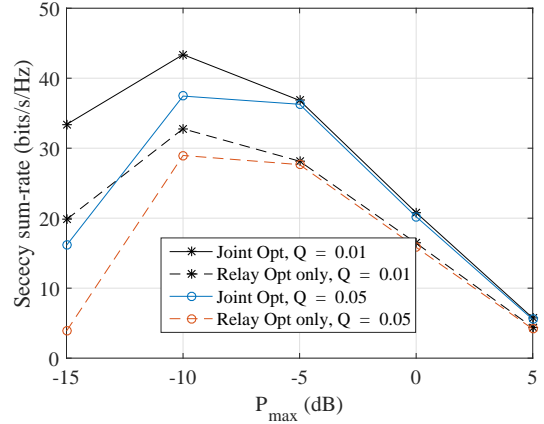}
\caption{Secrecy sum-rate versus $P_{\rm {max}}$.}\label{fig1}
\end{figure}

\begin{figure}[ht]
\centering
\includegraphics*[width=8cm]{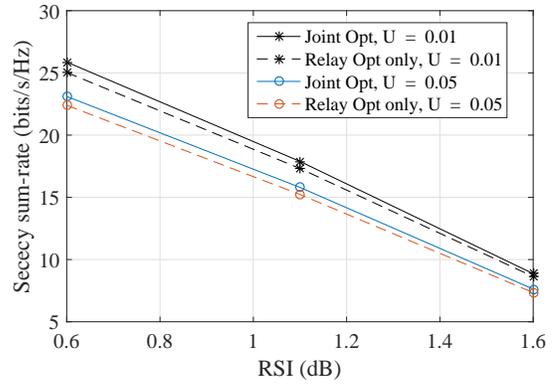}
\caption{Secrecy sum-rate versus RSI.}\label{fig2}
\end{figure}
In Fig. \ref{fig1}, we investigate the secrecy sum-rate for SWIPT in FD systems versus the transmit power budget $P_{\mathrm{max}}$ (dB) for different values of the harvested power constraint. We study the performance of the proposed scheme (denoted `Joint Opt.' in the figure) in comparison with the relay-only optimization scheme (denoted `Relay Only Opt.'). We see that the proposed scheme yields a higher secrecy sum-rate than the relay-only optimization scheme. Also, it can be observed that the secrecy sum-rate increases with $P_{\rm {max}}$ to a certain level after which it begins to experience a decrease with a continuous increase in $P_{\rm {max}}$, due to the increase of RSI \cite{networks}.

In Fig. \ref{fig2}, we investigate further the secrecy sum-rate performance against the RSI for different values of the harvested power constraints. Evidently, as RSI increases, a corresponding decrease in the secrecy sum-rate is observed. However, the proposed scheme yields higher secrecy sum-rate  compared to the secrecy sum-rate of the relay-only optimization scheme for different values of the harvested energy constraint. Hence, the need for joint optimization is justified.

\section{Conclusion}
This letter investigated the joint optimization of the source transmit power, AN covariance matrix, and the relay beamforming matrix for SWIPT in FD AF relaying system in the presence of an eavesdropper. Specifically, using SDP and 1-D searching, we proposed an algorithm that minimizes the total transmit power for secure SWIPT in a FD MIMO AF relay system. 

\end{document}